\documentclass{INTERSPEECH2023}
\usepackage{array} 
\usepackage{makecell}
\usepackage{diagbox}
\usepackage{float}
\usepackage{amsmath}
\usepackage{multirow}
\usepackage{pdfpages}
\usepackage{pifont}
\usepackage{setspace}
\usepackage{threeparttable}
\usepackage{CJKutf8}

\interspeechcameraready

\title{Unsupervised Active Learning: Optimizing Labeling Cost-Effectiveness for Automatic Speech Recognition}

\name{Zhisheng Zheng$^1$, Ziyang Ma$^1$, ‪Yu Wang$^{2,3}$, Xie Chen$^{1\ast}$\thanks{$\ast$  Corresponding author\\\indent \indent This work was supported by the National Natural Science Foundation of China (No. 62206171 and No.62106140), the International Cooperation Project of PCL, and Alibaba Group through Alibaba Innovative Research Program.}}
\address{
  $^1$MoE Key Lab of Artificial Intelligence, AI Institute, X-LANCE Lab \\ Department of Computer Science and Engineering, Shanghai Jiao Tong University\\
  $^2$Cooperative Medianet Innovation Center, Shanghai Jiao Tong University\\
  $^3$Shanghai AI Laboratory}
  
\email{\{zzs666, zym.22, yuwangsjtu, chenxie95\}@sjtu.edu.cn}

\begin{document}
\begin{CJK}{UTF8}{gbsn}

\maketitle


\begin{abstract}
In recent years, speech-based self-supervised learning (SSL) has made significant progress in various tasks, including automatic speech recognition (ASR). An ASR model with decent performance can be realized by fine-tuning an SSL model with a small fraction of labeled data. Reducing the demand for labeled data is always of great practical value. In this paper, we further extend the use of SSL to cut down labeling costs with active learning. Three types of units on different granularities are derived from speech signals in an unsupervised way, and their effects are compared by applying a contrastive data selection method. The experimental results show that our proposed data selection framework can effectively improve the word error rate (WER) by more than 11\% with the same amount of labeled data, or halve the labeling cost while maintaining the same WER, compared to random selection.


\end{abstract}

\noindent\textbf{Index Terms}: speech recognition, self-supervised learning, fine-tuning, unsupervised data selection

\section{Introduction}


Self-supervised learning (SSL) has emerged as a promising machine learning paradigm that allows us to learn more robust and distinctive features from unlabeled data, by leveraging inherent structure inside data without explicit labels. Recent works~\cite{hsu2021hubert,baevski2020wav2vec,baevski2022data2vec} have demonstrated that SSL models can extract high-quality and generalizable speech representations for various downstream speech-related tasks, such as speech recognition, speaker verification, and emotion recognition~\cite{yang21c_interspeech}. \\
\indent The SSL model typically consists of two-stage training: pre-training and fine-tuning. In the pre-training phase, several studies~\cite{lu22_interspeech,park2022unsupervised} have demonstrated that using higher quality unlabeled speech data can improve the model's generalization performance. On the other hand, to achieve high performance in downstream tasks like ASR, it is necessary to fine-tune the pre-trained model with task-specific labeled data. However, acquiring labeled data in the fine-tuning stage can be expensive and challenging. Therefore, a practical challenge is how to select domain-relevant or task-relevant speech data within a limited budget for annotation in order to maximize the cost-effectiveness of labeling, making SSL models more practical and accessible.

\begin{figure}[!htp]
\centering
\includegraphics[width=82mm]{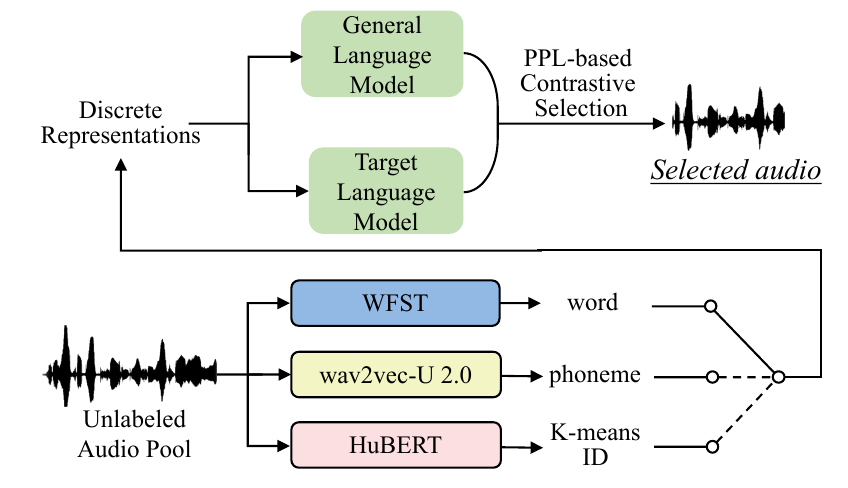}
\caption{The pipeline of unsupervised data selection on different granularities. The solid line represents we switch to word level. In this case the general language model is trained with word sequences from the WFST decoder, and the target language model is obtained by fine-tuning the general model with domain-specific text. Different granularities levels (K-means ID, phoneme, and word) are investigated. }
\label{Fig:select}
\vspace{-0.4cm}
\end{figure}

\indent In this paper, we present a completely unsupervised framework for selecting domain-relevant speech data. As illustrated in Fig.~\ref{Fig:select}, the process involves generating discrete token sequences at different levels of granularities (K-means ID, phoneme, word) from unlabeled speech data using intermediate models such as HuBERT, wav2vec-U 2.0, and WFST. We then calculate the perplexity (PPL) of these sequences using two pre-trained language models (a general LM and a target LM). Finally, we apply our PPL-based contrastive data selection approach to select the speech data that is most relevant to the target text.\\
\indent Similar to previous works~\cite{lu22_interspeech,chiu2022self}, we use the discrete representations as the input units for our language model. Baevski et al.~\cite{baevski2021unsupervised} pointed out that the discrete representations generated at different stages may capture different levels of acoustic information, which could potentially affect the accuracy of language modeling and the effectiveness of subsequent data selection. Therefore, in addition to the final recognition accuracy, other factors such as the amount of labeled data required for fine-tuning and the computational complexity of the process should also be considered when selecting the most appropriate granularities level for speech recognition. \\
\indent We demonstrate the efficacy of our proposed unsupervised data selection method in the fine-tuning stage on a subset of GigaSpeech~\cite{chen21o_interspeech}. At the same level of granularity, by selecting only 100 hours of speech audio that closely match the given corpus, and fine-tuning the HuBERT base model on this labeled data without using any language models, we are able to reduce the Word Error Rate (WER) by more than 11\% on all evaluated target domains. The main contributions can be summarized in three folds: 
\begin{itemize}
  \item We propose a novel, completely unsupervised active learning framework for speech data selection, which effectively reduces the cost of data labeling.
  \item We analyze the impact of different granularities levels on data selection and measure the trade-off between process complexity and recognition accuracy.
  \item Our proposed framework can either reduce the WER by over 11\% with the same amount of labeled data, or cut the labeling cost to half while maintaining the same WER, compared to random selection.
\end{itemize}

\section{Related work}
\subsection{Unsupervised data selection}
Unsupervised data selection is a crucial technique aimed at achieving the goal of reducing the need for labeled data while maintaining high performance in downstream tasks for a specified target domain. In the field of natural language processing (NLP), various approaches~\cite{ramponi-plank-2020-neural,blei2003latent,jelodar2019latent} have been proposed for unsupervised data selection, including domain adaptation and topic models. In the field of ASR, however, the discrete representations of speech are not explicit, which reduces the possibility of borrowing methods from the NLP field. Therefore, a key challenge is how to obtain discrete token representations from continuous speech signals for speech data slection. Lu et al.~\cite{lu22_interspeech} encoded speech signals into acoustically discrete tokens via self-supervised learning frameworks. Park et al.~\cite{park2022unsupervised} calculated frame-level losses on a target data set and a training data set through SSL models, and then averaged these losses at the utterance level for subsequent selection. In addition to SSL-based methods, traditional unsupervised methods are still applicable. Drugman et al.~\cite{drugman2019active} selectd data with low confidence scores from a speech recognition system. 
Malhotra et al.~\cite{malhotra2019active} proposed an entropy-based method for selecting the data that is most informative and uncertain for ASR.

\subsection{SSL models}
 Self-supervised learning has attracted a lot of attention in recent years due to its potential to overcome the limitations of supervised learning that require large amounts of labeled data. It can be viewed as a two-stage process: pre-training and fine-tuning. In the pre-training stage, a model is trained on a large amount of unlabelled data using diverse self-supervised criteria, such as generative~\cite{yue2021phonetically,liu2020masked,liu21l_interspeech}; 
contrastive~\cite{baevski2020wav2vec, oord2018representation, chung2021w2v}; predictive~\cite{hsu2021hubert,baevski2022data2vec,chen2022wavlm,ma2022mt4ssl}. These tasks help the model learn general representations from the data without human labels. 
In the fine-tuning stage, the pre-trained model is continuous to be trained using a smaller amount of labeled data on the target domain and the representations are transferred to adapt a specific downstream task, ultimately leading to improved performance.

\subsection{wav2vec-U 2.0}
Wav2vec-U 2.0~\cite{liu2023towards} is an enhanced ASR system with a simplified architecture that achieves higher accuracy without requiring any pre-processing on the audio-side. Similar to its predecessor~\cite{baevski2021unsupervised}, wav2vec-U 2.0 learns the structure of speech from unlabeled audio data via self-supervised speech representations derived from either wav2vec 2.0~\cite{baevski2020wav2vec} or XLSR~\cite{xu2021simple} models. These speech representations are then mapped to phonemes through a Generative Adversarial Network (GAN).


\section{Methods} 

\begin{figure}[!htp]
\centering
\includegraphics[width=65mm]{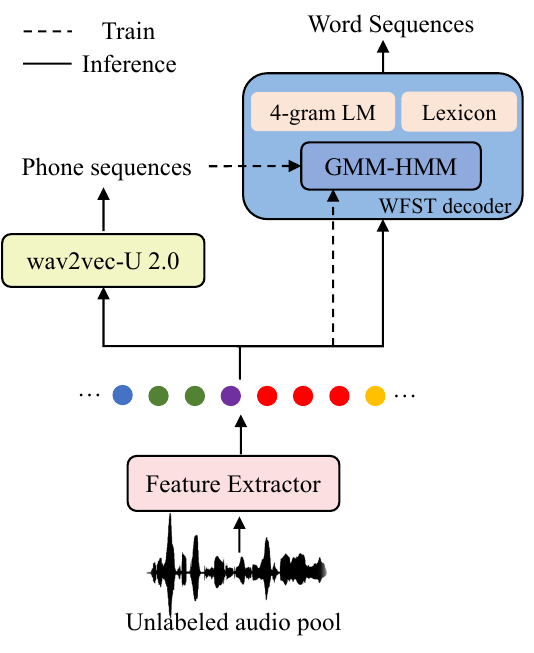}
\caption{The process of decoding audio into word sequences. Feature extractor is the pre-trained wav2Vec 2.0 Large (LV-60) model. 4-gram language model and lexicon are from the data preparation stage for training the wav2vec-U 2.0 model.}
\vspace{-0.45cm}
\label{Fig:2}
\end{figure}

In this section, we begin by addressing the process of obtaining phoneme labels from raw audio data in \S~\ref{subsec:w2vu}, followed by a description of our transducer that decodes the unlabeled audio into words in \S~\ref{HMM}. Both of these sections are illustrated in Fig.~\ref{Fig:2}. The unsupervised data selection strategy is elaborated upon in \S~\ref{LSTM}.

\subsection{Phoneme Recognizer}\label{subsec:w2vu}
Wav2vec-U 2.0 is a highly effective unsupervised ASR system that stands out for its ability to take raw representations extracted from SSL models such as wav2vec 2.0 or HuBERT as input and output the corresponding sequence of phonemes for a given speech signal.\\
\indent Prior to extracting features from speech using SSL models, it is necessary to perform VAD~\cite{tan2020rvad} on the speech signal to improve recognition accuracy. Subsequently, the SSL models like wav2vec 2.0 are used as feature extractors to obtain representations from the preprocessed speech. These extracted features are then fed into a pre-trained wav2vec-U 2.0 model, which outputs a sequence of phonemes as the final transcription.

\subsection{HMM-based Transcription}\label{HMM}

Section \S~\ref{subsec:w2vu} has outlined the process of obtaining high-quality phoneme transcriptions, which can be used as targets for training a sufficiently robust GMM-HMM model. Instead of using Mel Frequency Cepstral Coefficient (MFCC) feature as the input to the GMM-HMM model, we use the same frame-level representations from the feature extractor as described in \S~\ref{subsec:w2vu}. \\
\indent In order to generate word sequences, we train a 4-gram language model and create a lexicon from a public text corpus, which is also used for building the text input for the wav2vec-U 2.0 model. These are then utilized to construct a Weighted Finite State Transducer (WFST) system in combination with the previous trained HMM model. This system allows us to generate word pseudo-sequences from the unlabelled audio data.

\subsection{Contrastive data selection}\label{LSTM}
Contrastive data selection is a technique aimed at selecting samples from a larger dataset that are most relevant or similar to a target domain or task. Unlike \cite{lu22_interspeech}, where acoustically discrete labels are used as LM input, we utilize sub-word units (BPE) to train two language models (LM). The first LM is trained on sub-word level pseudo labels since they are more relevant to our task. Although we attempt to use publicly available text datasets as the training corpus for the model, the results are mediocre. The second language model is obtained by simply fine-tuning the first LM using a limited sample of domain-specific data. \\
\indent With these two trained LMs, we calculate the PPL of each sentence individually. However, our contrastive data selection algorithm does not directly use these sentence-level PPL, but instead uses the PPL of audio-level. 
Data selection at the utterance level may be more complex and inaccurate, and may require more prior knowledge because a single piece of text cannot fully represent the topic, sentiment, semantics, and other aspects of a speech, especially if the text is of low quality with high WER. Compared to that, audio-level selection method can mitigate the impact of irrelevant information, such as short sentences composed of common words and multi-topic words, thus allowing our algorithm to focus more on selecting topic-related audio.

The equation for the perplexity-based contrastive selection of each audio is defined as follows:
\begin{equation*}\label{eq:ppl}
    \eta = \frac{\overline{PPL}_{LM2} - \overline{PPL}_{LM1}}{\overline{PPL}_{LM1}}
\end{equation*}
where $\overline{PPL}$ denotes the average perplexity calculated for all utterances in every audio. 



Subsequently, we select the audio within a budget based on the ascending order of $\eta$. See Fig.~\ref{Fig:select} for an illustration.



\section{Experiments}
\subsection{Datasets}
\subsubsection{LibriSpeech and GigaSpeech}
The LibriSpeech corpus~\cite{panayotov2015librispeech} is a widely-used speech dataset that contains approximately 1,000 hours of transcribed audio data from read English audio books. The GigaSpeech corpus~\cite{chen21o_interspeech} is a large-scale multi-domain English dataset that consists of over 10,000 hours of high quality labeled audios, covering a diverse topics, such as \textit{Crime, Science, News}, etc. 

\subsubsection{Cross-Domain Dataset} 
This paper presents a medium-sized dataset, consisting of a 1,000-hour cross-domain subset of GigaSpeech. The dataset is unique in its multi-source, multi-style composition, with each theme comprising an equal amount of data.  \\
\indent The dataset we have compiled includes 4 topics, namely \textit{Crime, Health and Fitness, Howto and Style, and Science and Technology}. To ensure that the dataset is topic-balanced, we have included 100 hours of audio data for each of the four topics in the training set. To further augment the dataset and bring the total amount of data to 1000 hours, we have added an additional 600 hours of audio data from audiobooks, podcasts and youtube that are not specific to any of the four topics to the training set. In addition to the training set, we have constructed dedicated validation and test sets for each of the 4 topics. Each of these sets contains 5 hours of audio data, providing ample material for model development, evaluation, and comparison.  \\
\indent For the integrity and representativeness of the training, validation, and test sets, we have taken care to avoid any overlap in audio segments between these sets during the sampling process. Specifically, for the validation and test sets, we have sampled from the \textit{M}-size GigaSpeech training subset, as the word error rate of this subset is 0\%. For the training set, however, we require 100 hours of topic-specific data for each of the four topics, which cannot be fully provided by the \textit{M}-size and \textit{L}-size subsets. Therefore, we have sampled the remaining audio data from the \textit{XL}-size subset, resulting in approximately 270 hours of audio data with a word error rate of 4\%.

\subsection{Setup}
\begin{figure}[!htp]
\centering
\includegraphics[width=80mm]{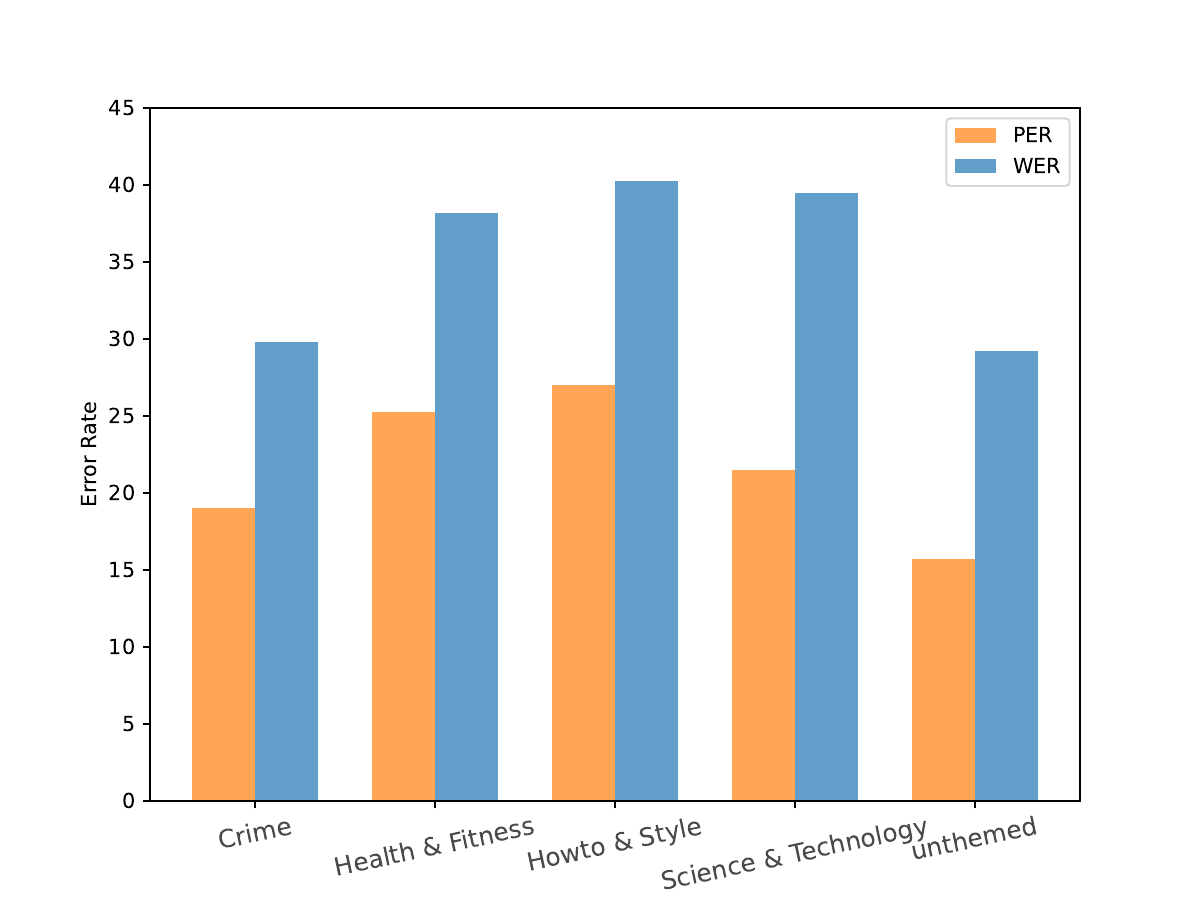}
\caption{Phone and word error rates on different categories. Phoneme recognition results are obtained from wav2vec-U 2.0 decoding, while word recognition results are from the decoding of the WFST decoder.}
\vspace{-0.3cm}
\label{Fig:uer}
\end{figure}

For our experiments, we use the pre-trained wav2Vec 2.0 Large (LV-60) model as the feature extractor to obtain speech representations. We use these representations extracted from 960 hours of LibriSpeech and 30,000 randomly selected text samples from a publicly available training corpus to train a sufficiently good GAN model in the wav2vec-U 2.0 system~\cite{liu2023towards}. The final model achieves a PER of 10.9\% on the LibriSpeech dev-other. We use this model to decode the training set of our Cross-Domain dataset and obtain phoneme pseudo-sequences, which have a PER of approximately 18.8\% (as shown in Fig.~\ref{Fig:uer}). \\
\indent To improve recognition accuracy, we train a GMM-HMM model with speech representations extracted from the SSL model as input and utilize the pseudo-labels as targets, resulting in a final PER of 15.4\%. We construct a WFST decoder by combining the GMM-HMM model, 4-gram language model, and lexicon. This allows us to decode the raw audio data to obtain word-level transcriptions, which results in a WER of around 32.4\% (as shown in Fig.~\ref{Fig:uer}).

As shown in Fig.~\ref{Fig:select}, we extract the representations from the sixth layer~\cite{hsu2021hubert} of the HuBERT base model to obtain K-Means IDs. We then apply the K-means algorithm to cluster the representations into 500 classes, and use the resulting K-means clustering IDs as direct input for the language models. For phoneme-based discrete representations, we also directly use them as inputs for the language model without any further processing. However, considering the scalability of the word vocabulary and the performance of our small model, we constructed a sub-word corpus with a vocabulary size of 5000 using the BPE algorithm. \\
\indent We then train our first language model using discrete token corpus, employing a Long Short-Term Memory (LSTM)~\cite{hochreiter1997long} with 2 layers of hidden units and a vector dimension of 768. Several key hyperparameters of this model are set as follows: the learning rate is 1.0e-4, the number of epochs is set to 10 and the dropout rate is 0.2. As for the second language model, we simply fine-tune the first one using the given domain text with the same lexicon. \\
\indent To evaluate the quality of our selected 100 hours of data, we employ the off-the-shelf pre-trained HuBERT base model as the quality assessor. We fine-tune each model for 80,000 steps and use the Viterbi algorithm as the decoding method.

\subsection{Results}

\begin{table}[htp]
\begin{center}
\caption{WERs of contrastive data selection on different granularities. All reported results are obtained by fine-tuning the HuBERT base model on 100 hours of labeled data and utilizing the Viterbi algorithm without language models. }
\label{tab:main}
\renewcommand{\arraystretch}{1.25} 
\begin{threeparttable}
\resizebox{1\columnwidth}{!}{
    \begin{tabular}{ccccc}
    \specialrule{0.10em}{0pt}{1pt}
     \multirow{2}{*}{\shortstack{Data Selection \\Algorithm}} & \multirow{2}{*}{Crime} & \multirow{2}{*}{\shortstack{Health \\and Fitness}} & \multirow{2}{*}{\shortstack{Howto \\and Style}} & \multirow{2}{*}{\shortstack{Science \\and Technology}} \\
     & & & &  \\
     \specialrule{0.10em}{0pt}{0pt}
    Random & 7.11 & 8.87 & 8.96 & 9.12 \\
    Categorized\tnote{1} & 6.04 & 7.58 & 8.16 & 7.91 \\
    \specialrule{0.07em}{0.5pt}{0.5pt}
    \multicolumn{4}{l}{\textbf{\quad PPL-based Contrastive Selection (Granularities)}} \\
    K-means id & 6.40 & 7.73 & 8.13 & 8.22 \\
    Phoneme & 7.39 & 8.85 & 8.3 & 8.63 \\
    Words & 6.26 & 7.57 & 7.96 &  8.12 \\
    Words*\tnote{2} & 6.05 & 7.65 & 7.97  & 8.01 \\
    \specialrule{0.10em}{0pt}{0pt}
    \end{tabular}
}
 \begin{tablenotes}
        \footnotesize
        \item[1] Data labeled with domain-specific classification tags.
        \item[2] Ground truth word sequences.
      \end{tablenotes}
\end{threeparttable}
\vspace{-0.5cm}
\end{center}
\end{table}

Table~\ref{tab:main} shows the WERs on the test set of our \textit{Cross-Domain} dataset, with different data selection strategies and discrete token granularities levels. All results are obtained from decoding the test set without any language models, using the same HuBERT base model fine-tuned on 100-hour labeled data. The first two rows show the results of a random selection of labeled data and data labeled with domain-specific classification tags, respectively. The following rows show the results of the PPL-based contrastive selection method with different discrete token granularities levels, including K-means id, phoneme, and word-level tokens. The last row shows the results of using ground truth word sequences as LM input. In general, the PPL-based contrastive selection method outperforms the random sampling in almost all cases, regardless of granularities levels. The use of word-level tokens yields the best results (relatively more than 11\%) across all domains, with ground truth word sequences performing even better, while phoneme-level tokens result in the highest WERs. 


\begin{table}[htp]
\begin{center}
\caption{WERs with different amounts of labeled data. Results are evaluated for labeled data of different durations, using the same data selection algorithm.}
\label{tab:time}
\renewcommand{\arraystretch}{1.25} 
\resizebox{1\columnwidth}{!}{
    \begin{tabular}{c|c|cccc}
    \specialrule{0.10em}{0pt}{1pt}
     \multirow{2}{*}{\shortstack{Data Selection \\Algorithm}} & \multirow{2}{*}{\shortstack{Labeled \\ data}} & \multirow{2}{*}{Crime} & \multirow{2}{*}{\shortstack{Health \\and Fitness}} & \multirow{2}{*}{\shortstack{Howto \\and Style}} & \multirow{2}{*}{\shortstack{Science \\and Technology}} \\
     & & & &  & \\
     \specialrule{0.10em}{0pt}{0pt}
    \multicolumn{1}{c|}{Random} & 100h & 7.11 & 8.87 & 8.96 & 9.12 \\
    \specialrule{0.07em}{0.5pt}{0.5pt}
    \multirow{4}{*}{Words} & 100h & 6.26 & 7.57 & 7.96 &  8.17 \\
    & 80h & 6.47 & 7.83 & 8.36 & 8.44 \\
    & 60h & 6.83 & 8.24 & 8.85 & 9.02 \\
    & 50h & 7.02 & 8.59 & 8.98 & 9.13 \\
    \specialrule{0.10em}{0pt}{0pt}
    \end{tabular}
}
\vspace{-0.5cm}
\end{center}
\end{table}

\indent Table~\ref{tab:time} compares the impact of labeled data of varying durations on the fine-tuning performance of SSL models, using the same word-level contrastive data selection method. Compared to a random sample of 100 hours of labeled data
 using our framework, we can achieve similar performance with only 50 hours of labeled data, which means the cost of labeling has been cut in half.

\subsection{Granularity Analysis}
Based on the results of Table~\ref{tab:main} and the difficulty in obtaining discrete representations, we can conduct a granularities analysis to determine which level of granularities is the most effective for speech recognition. Although using pseudo words as the granularities achieves the best performance and this performance can be further improved as the WER decreases, the process requires a multi-step inference procedure, which may be challenging to implement. In such situations, K-means ID-level representations may be a more practical alternative as they are the easiest to obtain. Compared with word sequences, HuBERT K-means IDs are derived from the acoustic features and may capture more detailed information about the acoustic characteristics of the speech signal. This may be the reason why using K-means IDs can achieve good performance.

\section{Discussion}
In this work, we made initial attempts to explore how to perform data selection based on completely unsupervised methods. Although the proposed algorithm has achieved good performance, there are several intriguing aspects that are worth investigating:
\begin{itemize}
    \item Is there an optimal level of granularities for discrete tokens in speech data selection?
    \item Can this data selection algorithm be applied to other speech-related downstream tasks beyond ASR?
    \item Is it possible to develop a more optimal and simpler unsupervised data selection strategy for active learning?
\end{itemize}
We will research these issues in the future.

\section{Conclusion}
In recent years, self-supervised learning (SSL) for speech has demonstrated promising results in enhancing different downstream tasks such as speech recognition. In this paper, we investigate the problem of reducing the labeling cost while maintaining high performance in ASR through efficient data selection for SSL fine-tuning within a limited budget. We present a fully unsupervised and flexible active learning framework that selects relevant data based on the perplexity-based contrastive selection method. We analyze and compare the effectiveness of our framework using three different levels of granularities for discrete tokens: K-means ID, phoneme, and word. The optimal level is determined based on the selection performance and the complexity of the process. Our experimental results confirm the effectiveness of our framework for SSL fine-tuning data selection, which achieves significant improvements in WER while being more cost-effective in terms of annotation.




\clearpage
\bibliographystyle{IEEEtran}
\bibliography{mybib}

\begin{thebibliography}{10}
\providecommand{\url}[1]{#1}
\csname url@samestyle\endcsname
\providecommand{\newblock}{\relax}
\providecommand{\bibinfo}[2]{#2}
\providecommand{\BIBentrySTDinterwordspacing}{\spaceskip=0pt\relax}
\providecommand{\BIBentryALTinterwordstretchfactor}{4}
\providecommand{\BIBentryALTinterwordspacing}{\spaceskip=\fontdimen2\font plus
\BIBentryALTinterwordstretchfactor\fontdimen3\font minus
  \fontdimen4\font\relax}
\providecommand{\BIBforeignlanguage}[2]{{%
\expandafter\ifx\csname l@#1\endcsname\relax
\typeout{** WARNING: IEEEtran.bst: No hyphenation pattern has been}%
\typeout{** loaded for the language `#1'. Using the pattern for}%
\typeout{** the default language instead.}%
\else
\language=\csname l@#1\endcsname
\fi
#2}}
\providecommand{\BIBdecl}{\relax}
\BIBdecl

\bibitem{hsu2021hubert}
W.-N. Hsu, B.~Bolte, Y.-H.~H. Tsai, K.~Lakhotia, R.~Salakhutdinov, and
  A.~Mohamed, ``{HuBERT}: Self-supervised speech representation learning by
  masked prediction of hidden units,'' in \emph{Proc. ICASSP 2021}, 2021, pp.
  3451--3460.

\bibitem{baevski2020wav2vec}
A.~Baevski, Y.~Zhou, A.~Mohamed, and M.~Auli, ``wav2vec 2.0: A framework for
  self-supervised learning of speech representations,'' \emph{Proc. NIPS 2020},
  pp. 12\,449--12\,460, 2020.

\bibitem{baevski2022data2vec}
A.~Baevski, W.-N. Hsu, Q.~Xu, A.~Babu, J.~Gu, and M.~Auli, ``Data2vec: A
  general framework for self-supervised learning in speech, vision and
  language,'' in \emph{Proc. ICML 2022}, 2022, pp. 1298--1312.

\bibitem{yang21c_interspeech}
S.~wen Yang, P.-H. Chi, Y.-S. Chuang, C.-I.~J. Lai, K.~Lakhotia, Y.~Y. Lin,
  A.~T. Liu, J.~Shi, X.~Chang, G.-T. Lin, T.-H. Huang, W.-C. Tseng, K.~tik Lee,
  D.-R. Liu, Z.~Huang, S.~Dong, S.-W. Li, S.~Watanabe, A.~Mohamed, and
  H.~yi~Lee, ``{SUPERB: Speech Processing Universal PERformance Benchmark},''
  in \emph{Proc. Interspeech 2021}, 2021, pp. 1194--1198.

\bibitem{lu22_interspeech}
Z.~Lu, Y.~Wang, Y.~Zhang, W.~Han, Z.~Chen, and P.~Haghani, ``{Unsupervised data
  selection via discrete speech representation for ASR},'' in \emph{Proc.
  Interspeech 2022}, 2022, pp. 3393--3397.

\bibitem{park2022unsupervised}
C.~Park, R.~Ahmad, and T.~Hain, ``Unsupervised data selection for speech
  recognition with contrastive loss ratios,'' in \emph{Proc. ICASSP 2022},
  2022, pp. 8587--8591.

\bibitem{chiu2022self}
C.-C. Chiu, J.~Qin, Y.~Zhang, J.~Yu, and Y.~Wu, ``Self-supervised learning with
  random-projection quantizer for speech recognition,'' in \emph{Proc. ICML
  2022}, 2022, pp. 3915--3924.

\bibitem{baevski2021unsupervised}
A.~Baevski, W.-N. Hsu, A.~Conneau, and M.~Auli, ``Unsupervised speech
  recognition,'' \emph{Proc. NIPS 2021}, pp. 27\,826--27\,839, 2021.

\bibitem{chen21o_interspeech}
G.~Chen, S.~Chai, G.-B. Wang, J.~Du, W.-Q. Zhang, C.~Weng, D.~Su, D.~Povey,
  J.~Trmal, J.~Zhang, M.~Jin, S.~Khudanpur, S.~Watanabe, S.~Zhao, W.~Zou,
  X.~Li, X.~Yao, Y.~Wang, Z.~You, and Z.~Yan, ``{GigaSpeech: An evolving,
  multi-Domain {ASR} corpus with 10,000 hours of transcribed audio},'' in
  \emph{Proc. Interspeech 2021}, 2021, pp. 3670--3674.

\bibitem{ramponi-plank-2020-neural}
A.~Ramponi and B.~Plank, ``Neural unsupervised domain adaptation in
  {NLP}{---}{A} survey,'' in \emph{Proc. ICCL 2020}, 2020, pp. 6838--6855.

\bibitem{blei2003latent}
D.~M. Blei, A.~Y. Ng, and M.~I. Jordan, ``Latent {D}irichlet allocation,''
  \emph{Proc. JMLR 2003}, pp. 993--1022, 2003.

\bibitem{jelodar2019latent}
H.~Jelodar, Y.~Wang, C.~Yuan, X.~Feng, X.~Jiang, Y.~Li, and L.~Zhao, ``Latent
  {D}irichlet allocation ({LDA}) and topic modeling: models, applications, a
  survey,'' \emph{Proc. MTAs 2019}, pp. 15\,169--15\,211, 2019.

\bibitem{drugman2019active}
T.~Drugman, J.~Pylkkonen, and R.~Kneser, ``Active and semi-supervised learning
  in {ASR}: Benefits on the acoustic and language models,'' \emph{arXiv
  preprint arXiv:1903.02852}, 2019.

\bibitem{malhotra2019active}
K.~Malhotra, S.~Bansal, and S.~Ganapathy, ``Active learning methods for low
  resource end-to-end speech recognition.'' in \emph{Proc. Interspeech 2019},
  2019, pp. 2215--2219.

\bibitem{yue2021phonetically}
X.~Yue and H.~Li, ``Phonetically motivated self-supervised speech
  representation learning.'' in \emph{Proc. Interspeech 2021}, 2021, pp.
  746--750.

\bibitem{liu2020masked}
L.~Liu and Y.~Huang, ``Masked pre-trained encoder base on joint
  {CTC}-{Transformer},'' \emph{arXiv preprint arXiv:2005.11978}, 2020.

\bibitem{liu21l_interspeech}
A.~H. Liu, Y.-A. Chung, and J.~Glass, ``{Non-autoregressive predictive coding
  for learning speech representations from local dependencies},'' in
  \emph{Proc. Interspeech 2021}, 2021, pp. 3730--3734.

\bibitem{oord2018representation}
A.~v.~d. Oord, Y.~Li, and O.~Vinyals, ``Representation learning with
  contrastive predictive coding,'' \emph{arXiv preprint arXiv:1807.03748},
  2018.

\bibitem{chung2021w2v}
Y.-A. Chung, Y.~Zhang, W.~Han, C.-C. Chiu, J.~Qin, R.~Pang, and Y.~Wu,
  ``W2v-bert: Combining contrastive learning and masked language modeling for
  self-supervised speech pre-training,'' in \emph{Proc. ASRU 2021}, 2021, pp.
  244--250.

\bibitem{chen2022wavlm}
S.~Chen, C.~Wang, Z.~Chen, Y.~Wu, S.~Liu, Z.~Chen, J.~Li, N.~Kanda,
  T.~Yoshioka, X.~Xiao \emph{et~al.}, ``Wavlm: Large-scale self-supervised
  pre-training for full stack speech processing,'' \emph{Proc. IEEE J Sel Top
  Signal Process 2022}, pp. 1505--1518, 2022.

\bibitem{ma2022mt4ssl}
Z.~Ma, Z.~Zheng, C.~Tang, Y.~Wang, and X.~Chen, ``{MT4SSL}: Boosting
  self-supervised speech representation learning by integrating multiple
  targets,'' \emph{arXiv preprint arXiv:2211.07321}, 2022.

\bibitem{liu2023towards}
A.~H. Liu, W.-N. Hsu, M.~Auli, and A.~Baevski, ``Towards end-to-end
  unsupervised speech recognition,'' in \emph{Proc. SLT 2022}, 2022, pp.
  221--228.

\bibitem{xu2021simple}
Q.~Xu, A.~Baevski, and M.~Auli, ``Simple and effective zero-shot cross-lingual
  phoneme recognition,'' \emph{arXiv preprint arXiv:2109.11680}, 2021.

\bibitem{tan2020rvad}
Z.-H. Tan, A.~Kr.~Sarkar, and N.~Dehak, ``r{VAD}: An unsupervised segment-based
  robust voice activity detection method,'' \emph{Proc. Computer speech \&
  language 2020}, pp. 1--21, 2020.

\bibitem{panayotov2015librispeech}
V.~Panayotov, G.~Chen, D.~Povey, and S.~Khudanpur, ``Librispeech: An {ASR}
  corpus based on public domain audio books,'' in \emph{Proc. ICASSP 2015},
  2015, pp. 5206--5210.

\bibitem{hochreiter1997long}
S.~Hochreiter and J.~Schmidhuber, ``Long short-term memory,'' \emph{Proc.
  Neural computation 1997}, pp. 1735--1780, 1997.

\end{thebibliography}

\end{CJK}
\end{document}